\documentclass[letterpaper, 10 pt, conference]{ieeeconf}  

\IEEEoverridecommandlockouts                              
\usepackage{epsfig} 
\usepackage{amsmath} 
\usepackage{amssymb}  
\usepackage{cite}

\title{\LARGE \bf
Robustness of force and stress inference in an epithelial tissue
}

\author{Kaoru Sugimura$^{1}$, Yohanns Bella\"{i}che$^{2}$, Fran\c{c}ois Graner$^{2, 3}$, Philippe Marcq$^{4}$, and Shuji Ishihara$^{5}$%
\thanks{*This work was supported by grants from MEXT (K.S. and S.I.) and PRESTO JST (S.I.). } 
\thanks{$^{1}$K. Sugimura is with Institute for Integrated Cell-Material Sciences (WPI-iCeMS), Kyoto University, Japan (ksugimura at 
icems.kyoto-u.ac.jp). $^{2}$Y. Bella\"{i}che and F. Graner 
are with Genetics and Developmental Biology, Team ``Polarity, division and morphogenesis,'' Institut Curie, France (yohanns.bellaiche at 
curie.fr). $^{3}$F. Graner is with Laboratoire Mati\`ere 
et Syst\`emes Complexes, Universit\'e Paris Diderot, Paris VII, France (francois.graner at univ-paris-diderot.fr). $^{4}$P. Marcq is with 
Physico-Chimie Curie, Institut Curie/CNRS-UMR168/
Universit\'e Pierre et Marie Curie, France (philippe.marcq at curie.fr). $^{5}$S. Ishihara is with Graduate School of Arts and Sciences, The 
University of Tokyo, Japan (shuji at complex.c.u-
tokyo.ac.jp). Corresponding author: Kaoru Sugimura
}
}

\begin{document}

\maketitle
\thispagestyle{empty}
\pagestyle{empty}

%
%

\begin{abstract}
During morphogenesis, the shape of a tissue emerges from collective cellular behaviors, which are in part regulated by mechanical and biochemical interactions 
between cells. Quantification of force and stress is therefore necessary to analyze the mechanisms controlling tissue morphogenesis. Recently, a mechanical 
measurement method based on force inference from cell shapes and connectivity has been developed. It is non-invasive, and can provide space-time maps of 
force and stress within an epithelial tissue, up to prefactors. We previously performed a comparative study of three force-inference methods, which differ in their 
approach of treating indefiniteness in an inverse problem between cell shapes and forces. In the present study, to further validate and compare the three force 
inference methods, we tested their robustness by measuring temporal fluctuation of estimated forces. Quantitative data of cell-level dynamics in a developing 
tissue suggests that variation of forces and stress will remain small within a short period of time ($\sim$minutes). Here, we showed that cell-junction tensions and 
global stress inferred by the Bayesian force inference method varied less with time than those inferred by the method that estimates only tension. In contrast, the 
amplitude of temporal fluctuations of estimated cell pressures differs less between different methods. Altogether, the present study strengthens the validity and 
robustness of the Bayesian force-inference method. 
\end{abstract}

%
%

\section{INTRODUCTION}
Epithelial tissue morphogenesis is regulated in part by forces acting along the plane of the adherens junction, \textit{i.e.}, tension that shortens a cell contact surface 
and pressure that counteracts the tension to maintain the size of a cell (Fig. \ref{fig:fig1}(a),(b)) \cite{Lecuit2011, Kasza2011, 
Graner1993,Honda2008,Rauzi2008,Aigouy2010,Bosveld2012}. Space-time maps of cell-junction tension, cell pressure, and tissue stress are therefore among the key 
aspects of physical information required to understand biomechanical control of morphogenesis. \textit{In vivo} mechanical measurement methods have already been 
reported \cite{Hutson2003,Bonnet2012,Nienhaus2009,Desprat2008}. Recently, methods based on force inference 
\cite{Stein1982,Brodland2010,Chiou2012,Ishihara2012} have been developed, which offer cell-level resolution, and are both non-invasive and global. They rely upon 
segmented images, \emph{i.e.} images wherein the cell contours and vertices have been recognized. Deviations from $120^{\circ}$ angles between cell 
contact surfaces indicate heterogeneities in tensions and pressures, which can be determined by solving a linear inverse problem (Sect. \ref{sec:FI}). Hence, forces 
within more than hundreds of cells can be simultaneously estimated. 

We previously performed a comparative study of three force-inference methods, which differ in their treatment of the indefiniteness in an inverse problem between cell 
shapes and forces (Sect. \ref{sec:FI}) \cite{Ishihara2013}. The first method (ST) estimates only tensions, and all the cell pressures are assumed to be the same. The 
second method (SP) estimates only cell pressures under the assumption of uniform tensions. Such assumptions decrease the number of unknowns (cell junction 
tensions or cell pressures) so that the first two methods treat overdetermined problems. The third method (STP) treats the ill-conditioned problem and simultaneously 
estimates both tensions and pressures by employing Bayesian statistics with a prior function representing positive tensions \cite{Ishihara2012}. Our results using 
different datasets consistently indicate that the Bayesian force inference (STP) performs best in terms of accuracy and robustness against image processing error 
\cite{Ishihara2013}. In the  present study, we performed another comparative test by measuring temporal fluctuation of forces in the \textit{Drosophila} wing.   Based 
on the results of this study, we will discuss the respective robustness of the three force-inference methods.

%
%

\section{METHODS}\label{methods}
\subsection{Image acquisition and analysis}
\begin{figure}[thpb]
      \centering
      \includegraphics[scale=0.53]{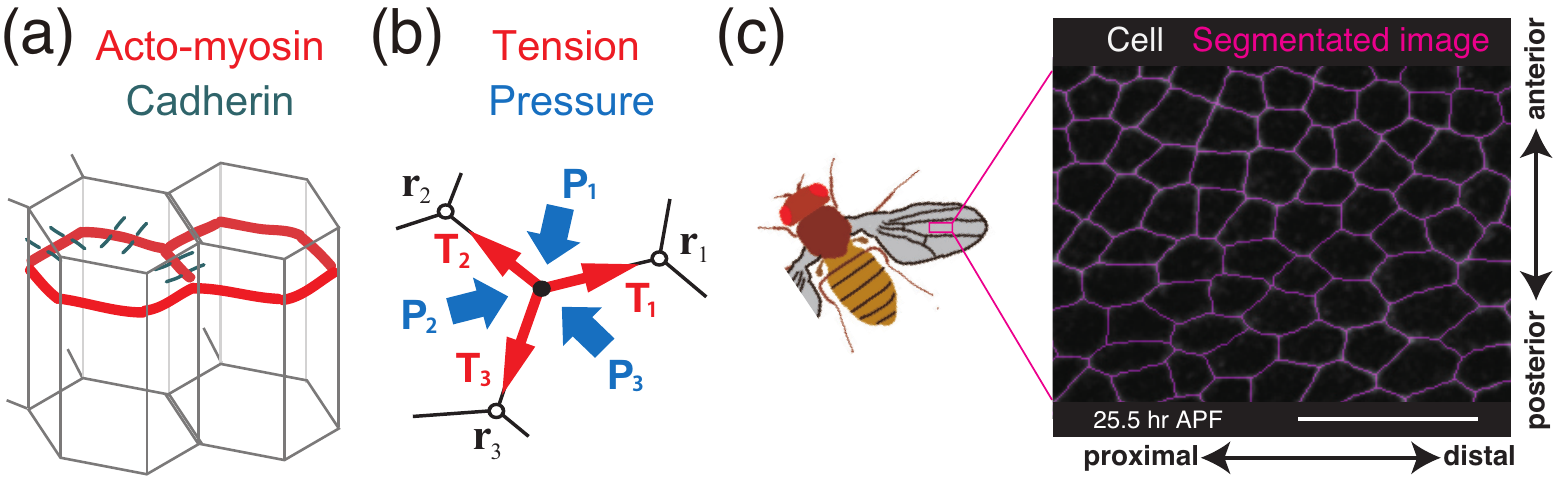}
      \caption{The structure and force balance of an epithelial tissue. (a) Mechanical interactions among epithelial cells act in the plane of the 
adherens junction, where cell adhesion molecules, cadherin, held cells together. Inside the cell, acto-myosin runs along the cell cortex in the plane of the 
adherens junction. (b) Forces acting on a vertex located at position $\vec{r}_0$ indicated by a black dot.  Tensions along the cell contact surfaces pull the vertex in the respective directions 
indicated by red arrows, while the cell  pressures push the vertex in the directions indicated by blue arrows. (c)  An image of a \textit{Drosophila} wing expressing DECad-GFP at 25.5 hr after 
puparium formation (APF). A segmented image (magenta) is overlapped. Scale bar: $20\;\mu \text{m}$.  }
      \label{fig:fig1}
\end{figure}

Preparation of samples of \textit{Drosophila} pupal wing for image collection was conducted as previously described in \cite{Ishihara2012}. 
Experiments were carried at 25.5 hr  and 31 hr after puparium formation (APF). To highlight the shape of the cell at the level of the adherens junction, DECadherin (DECad)-GFP 
\cite{Huang2009} was used (Fig. \ref{fig:fig1}(c)). Images (512 x 512 pixels; 0.188 pixel/$\mu \text{m}$) were acquired at 30 second interval for 10 minutes at 25$^
\circ$C using an inverted confocal microscope (A1R; Nikon) equipped with an 60x/NA1.2 Plan Apochromat water-immersion objective. We segmented images by 
using custom-made macros and plug-ins in ImageJ. We manually corrected the skeletonized pattern when necessary. A sample shown in Fig. \ref{fig:fig2} contains 
264 cells (62--64 outer cells and 200--202 inner cells) and 724--727 edges. 
 
\subsection{Outline of force estimation methods to be tested}\label{sec:FI}
Here, we briefly outline how one infers forces and stress from patterns of epithelial cell shapes and their connectivity, as already reported in detail in \cite{Chiou2012,Ishihara2012,Ishihara2013}. 

An epithelial cell sheet is approximated by a two-dimensional polygonal tile, and pressures of individual cells and tensions on the cell contact surfaces are assigned as unknown variables to be 
inferred, as $\textbf{P} = (P_1, \cdots, P_N)$ and $\textbf{T} = (T_1,\cdots,T_M)$ (Fig. \ref{fig:fig1}(b); $N$ and $M$ are numbers of cells 
and cell contact surfaces in the provided image). At each vertex, force balance is a linear equation 
\begin{eqnarray}
    \label{eq:1}
    A \textbf{X} = 0
\end{eqnarray}
with the unknown variables $\textbf{X} = (\textbf{T}, \textbf{P})$. In the equation, $A$ is an $n \times m$ matrix determined only from the observed geometry of the cell, where $n$ and $m$ 
signify the number of balance equations and unknown variables, respectively.  Cells are assumed to change their shapes quasi-statically, 
and possible drag forces are ignored. In the method, forces are estimated up to a scaling factor (they are normalized so that the mean value of tensions is unity). This problem is underdetermined 
($n<m$). Since force balance equations are invariant under a variation of (constant) hydrostatic pressure, the method infers only the difference of pressures ($\Delta \textbf{P}$) among cells. In 
what follows, we thus redefine $\textbf{X}$ as $ = (\textbf{T}, \Delta \textbf{P})$. Indefiniteness results from boundary conditions and from the presence of four-fold vertices, and can be managed 
in the general framework of inverse problem. The most likely solution $\textbf{X}$ can be uniquely determined from Eq. \ref{eq:1} in conjunction with prior knowledge of mechanical properties of 
the system. As described below, there are several possible ways to manage the indefiniteness that should be comparatively evaluated.

We tested three types of force-inference methods \cite{Ishihara2013}. We have called them ST, SP and STP, where the ``S" stands for  ``straight" edges (curvatures are neglected and cells are 
treated as polygons); ``T" and ``P" mean that tensions and pressures are unknown, respectively.

\textit{ST}: The first method ST assumes that the difference in pressures among cells is negligible ($\Delta P_i=0$), thus estimates only tensions ($\textbf{T}$) \cite{Chiou2012}.

\textit{SP}: The second method SP assumes that all tensions are uniform \emph{i.e.}, $T_{j} =1$. SP estimates only the difference of pressures among cells ($\Delta \textbf{P}$).

\textit{STP}: The third method STP is Bayesian Maximum A Posteriori (MAP) inference \cite{Kaipio2004}. Since laser severing experiments indicate that tensions are usually constricting in 
epithelial tissues \cite{Rauzi2008,Hutson2003}, we use as a prior a Gaussian distribution of tensions $T_{j}$ around a positive value \cite{Ishihara2012}. Briefly, pressures and tensions are 
determined by minimizing the function
\begin{eqnarray}
    \label{eq:2}
    S(\textbf{X}) = |A\textbf{X}|^2  + \mu  \sum {}_j (T_j-T_0)^2
\end{eqnarray}
with $T_0 = 1$. With use of Bayesian statistics, the weight of the second term, $\mu$, can be objectively determined by maximizing marginal likelihood function 
\cite{Akaike1980,Kaipio2004}.

\subsection{Global stress}\label{sec:GS}
Summing the estimates of $\mathbf{T}$ and $\Delta \mathbf{P}$ over the epithelium, one can deduce global stress according to Batchelor's formula 
\cite{Ishihara2012,Batchelor1970}:
\begin{eqnarray}
  \label{eq:StressTensor}
  \mathbf{\sigma} = \frac{1}{A}\left( -\sum_i P_i A_i {\mathbf I}+ \sum_j T_{j} \frac{\mathbf{r}_{j}\otimes\mathbf{r}_j}{|\bf{r}_j|} \right),
\end{eqnarray}where ${\mathbf I}$ is the two-dimensional identity tensor, the vectors $\mathbf{r}_j$ span the $j$-th cell edge, and $A 
\equiv \sum_i A_i$ is the total area of the epithelial domain.

%
%

\section{RESULTS}
The robustness of force-inference methods may be tested by several procedures. In \cite{Ishihara2012,Ishihara2013}, we performed the following two tests, at a given fixed time: \emph{(i)} 
Randomly delete small fraction of edges from a segmented image and measure the variance of global tissue stress; \emph{(ii)} 
Add noise to all extracted positions of vertices and measure the variance of cell junction tension, cell pressure, and global tissue stress. Here, we quantified the fluctuations of estimated forces 
and stress in time. If the time interval between successive images is short compared with the time scale of cell-level morphogenetic processes  such as cell growth, cell division, and cell 
rearrangement, variation of actual forces and stress are expected to be small.

To perform this latter test, we used images of \textit{Drosophila} pupal wing. During pupal development, wing cells undergo cell divisions, and the initial, nearly isotropic morphology of wing cells 
becomes elongated (15--24 hr APF). After that, the bias in the lengths of the edges exhibits a moderate decrease and the fraction of 
hexagonal cells increases through cell rearrangements (24--32 hr APF).  We analyzed images that were acquired at 30 second interval for 10 minutes at 25.5 hr and 31 hr APF. Quantitative data 
of cell-level dynamics in the wing suggests that forces change significantly over hours\cite{Aigouy2010}. We expect that average forces barely change over time-scale of minutes.

\begin{figure}[tb]
      \centering
      \includegraphics[scale=0.5]{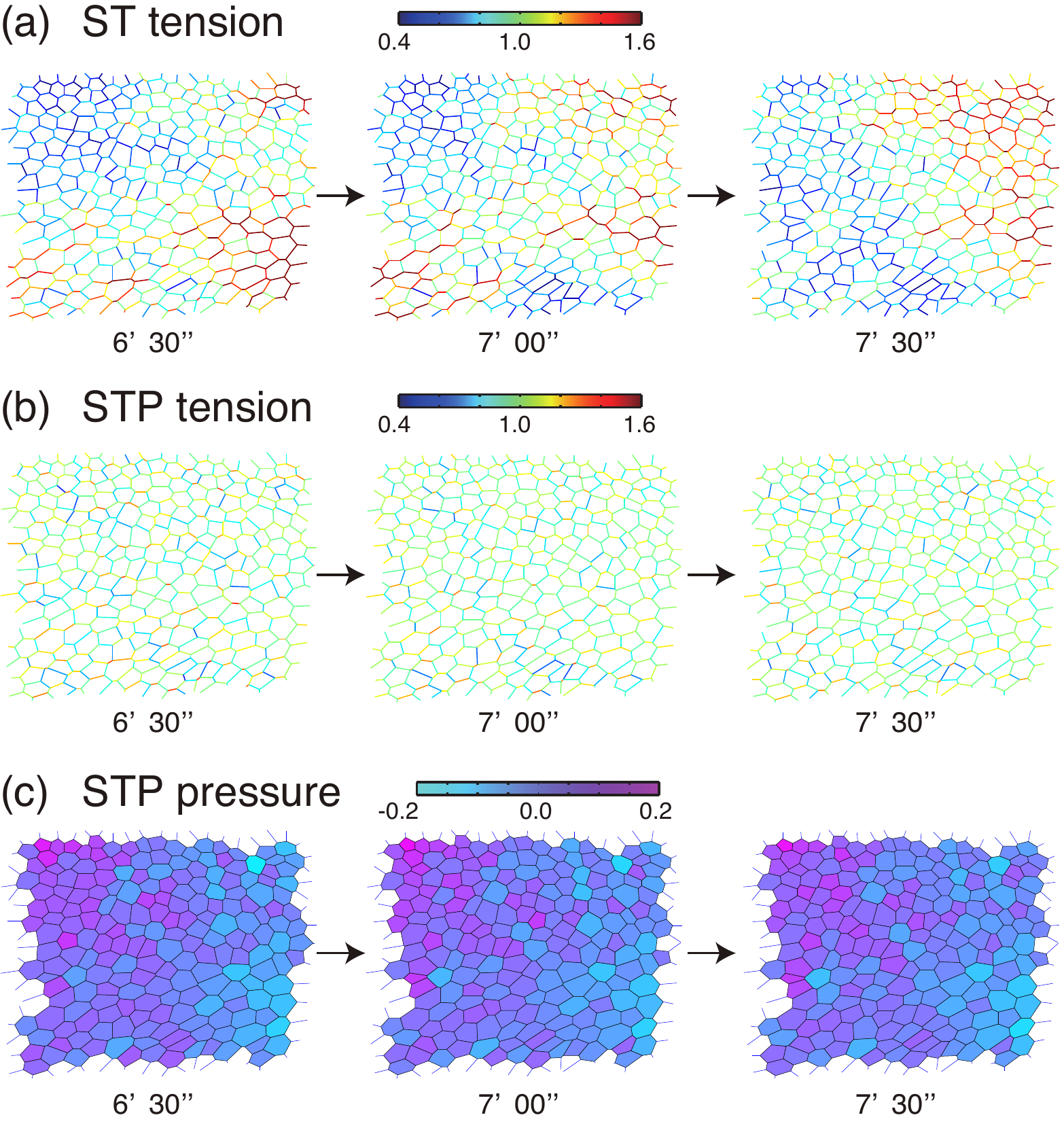}
      \caption{Maps of tensions and pressures of the \textit{Drosophila} wing at 25.5 hr APF. The vertical and horizontal directions are aligned with the anterior-
posterior and proximal-distal axes, respectively. (a)--(c) Panels correspond to three successive images (times indicated below the panels) and units of forces are 
adimensioned. (a) Tensions estimated using ST. (b) Tensions estimated using STP. (c) Pressures estimated using STP.}
      \label{fig:fig2}
\end{figure}

\begin{figure}[thpb]
      \centering
      Ê\includegraphics[scale=0.5]{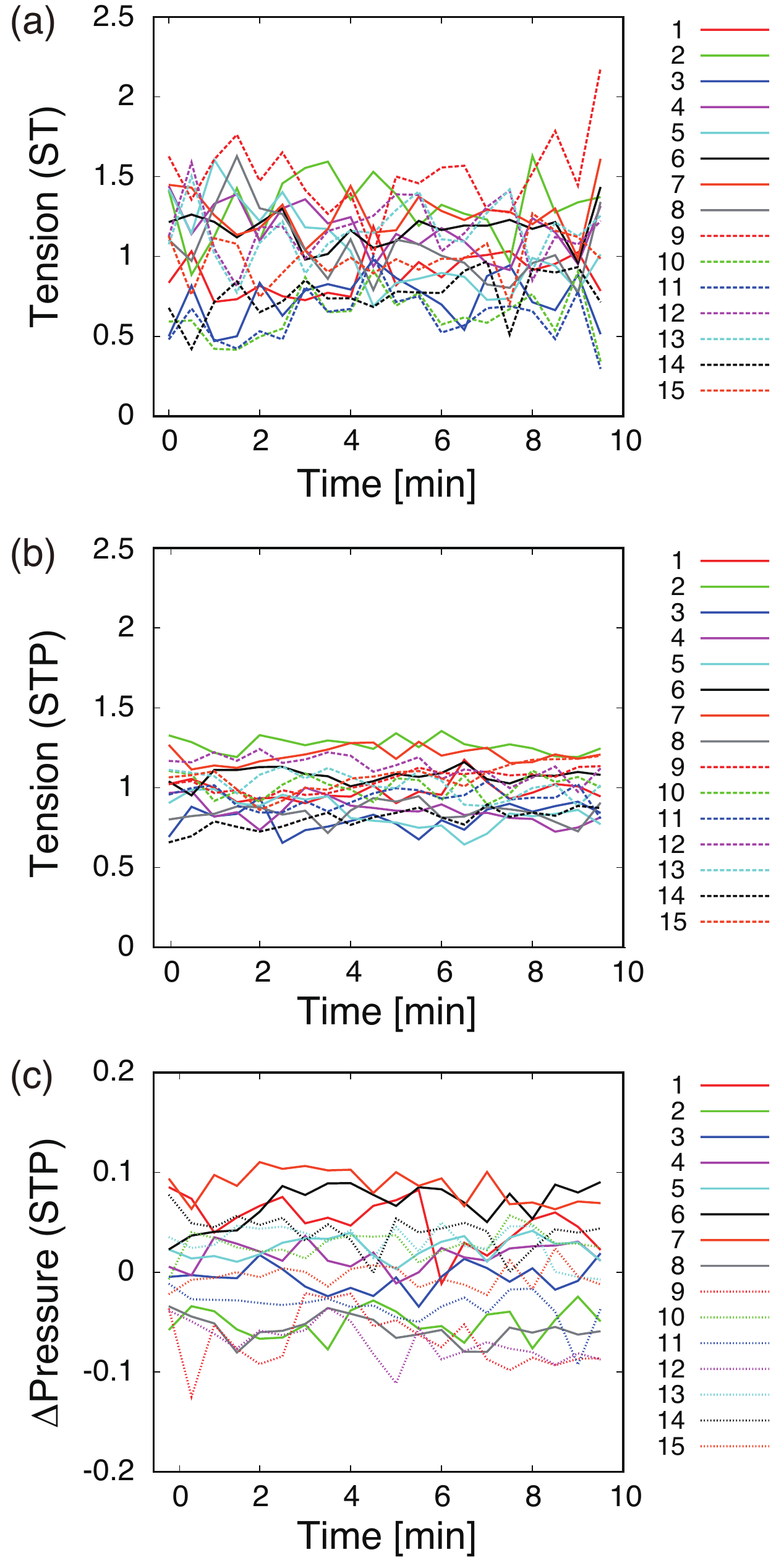}
      \caption{Single edge/cell tracking of estimated tensions and pressures. (a) Tensions estimated using ST. (b) Tensions estimated using STP. In (a) and (b), the same set of edges is tracked. (c) 
Pressures estimated using STP.}
      \label{fig:fig3}
\end{figure}

Fig. \ref{fig:fig2}(a) and (b) show the maps of estimated tensions in the wing at 25.5 hr APF, where tension values are indicated by a color scale (movies files can be downloaded from http://
koolau.info/movies/EMBC2013.zip). As reported in \cite{Ishihara2013}, maps of tensions obtained by ST showed ``patches" (distinct regions where the tension seems locally uniform), and the 
position of patches varies among successive images (Fig. \ref{fig:fig2}(a)). In contrast, maps of tensions estimated by using STP were relatively constant with time (Fig. \ref{fig:fig2}(b)). Single 
edge tracking analysis confirmed that tensions estimated by STP fluctuated less than those estimated by ST (Fig. \ref{fig:fig3}(a) and (b); see large jumps and drops in Fig. \ref{fig:fig3}(a)).    

The difference of pressures among cells obtained using STP is color-coded in Fig. \ref{fig:fig2}(c). The pressure map obtained using SP was similar to that obtained using STP (not shown) 
\cite{Ishihara2013}. As clearly seen from Fig. \ref{fig:fig3}(c), there is no drift of pressures in time. The standard deviation of $\Delta \textbf{P}$ among 20 time points for data shown in Fig. 
\ref{fig:fig3}(c) were 0.011 in SP, and 0.017 in STP. These values were smaller than the dispersion of the data at each time point (0.045 in SP and 0.05 in STP for all inner cells).

\begin{figure}[hpb]
      \centering
      Ê\includegraphics[scale=0.65]{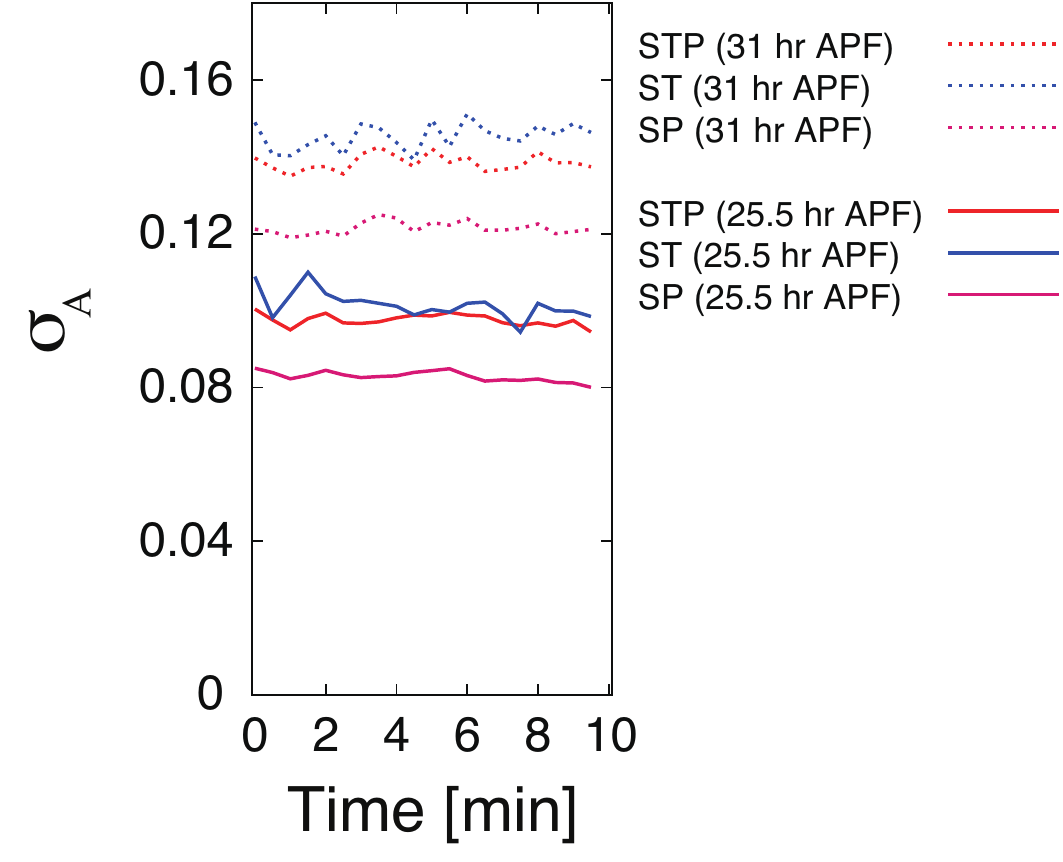}
      \caption{Temporal fluctuation of global stress of the wing. $\sigma_{A} \equiv (\sigma_{yy}-\sigma_{xx})$ estimated by STP (red), ST (blue), 
and SP (magenta) are plotted against time. Solid lines (25.5 hr APF) and dotted lines (31 hr APF). $x$ and $y$ axes correspond to the anterior-posterior and proximal-distal axes of the wing, 
respectively. }
      \label{fig:fig4}
\end{figure}

Finally, we evaluated global stress, which can be deduced by integrating tensions and pressures (Eq. \ref{eq:StressTensor}). The normal stress difference $\sigma_{A} \equiv (\sigma_{yy}-
\sigma_{xx})$ was used previously \cite{Ishihara2013} to cross-validate the Bayesian force inference with large-scale tissue ablation \cite{Bonnet2012}. It is independent of the (undetermined) 
value of the hydrostatic pressure, and characterizes the anisotropy of stress. We found that the deviation of $\sigma_A$ was larger in ST than in SP and STP (Fig. \ref{fig:fig4}). The standard 
deviations of $\sigma_A$ among 20 time points were $3.5 \times 10^{-3}$ in ST, $1.3\times 10^{-3}$ in SP, and $1.5 \times 10^{-3}$ in STP in time sequence of images of 25.5 hr APF (solid lines 
in Fig. \ref{fig:fig4}). These values were $3.5 \times 10^{-3}$ in ST, $1.6 \times 10^{-3}$ in SP, and $2.1 \times 10^{-3}$ in STP in data of 31 hr APF (dotted lines in Fig. \ref{fig:fig4}).

%
%

\section{DISCUSSION}\label{discussion}
In the present study, we quantified the temporal fluctuations of forces and stress in order to test the  robustness of three force-inference methods (ST, SP, and STP).
Our data indicated that within the time scale of movies analyzed, tensions and global stress obtained using STP vary less than those obtained by using ST. The appearance of ``patches" in a 
map of tensions (\textit{i.e.,} long-wavelength mode) resulted in the large fluctuation in ST. In contrast, the prior in STP works as the ``regularization term' to avoid over-fitting by eliminating the 
long-wavelength mode, which makes STP more robust (see Discussion of \cite{Ishihara2013}). Somewhat counter-intuitively, the solution of the underdetermined method (STP) is more robust 
than that of the overdetermined one (ST).

STP and ST agree in the determination of stress in Fig. \ref{fig:fig4}, STP being more robust. Then SP may underestimate it. This is because stronger tensions on the proximal-distal edges of the 
wing \cite{Aigouy2010,Ishihara2012} were not incorporated in SP.  

In conclusion, the results of this and our recent work \cite{Ishihara2013} showed that robustness and accuracy of estimation are optimal in the Bayesian force 
inference method (STP).


%
%

\section*{ACKNOWLEDGMENT}
We thank Boris Guirao for initial input to this study. K.S. is grateful to Yang Hong for DECad-GFP fly, Yuri Tsukahara for technical assistance, Minako Izutsu for fly foods, and the iCeMS imaging 
center for imaging equipment.

%
%

\addtolength{\textheight}{-12cm}   

\end{document}